\documentclass[a4paper,fleqn,usenatbib]{mnras}

\usepackage{newtxtext,newtxmath}

\usepackage[T1]{fontenc}
\usepackage{ae,aecompl}

\newcommand{\kms}{km\,s$^{-1}$}
\newcommand{\vs}{$v_{\rm e} \sin i$}
\newcommand{\teff}{$T_{\rm eff}$}
\newcommand{\lgg}{$\log\,{g}$}

\usepackage{graphicx}	
\usepackage{amsmath}	
\usepackage{amssymb}	

\title[Magnetometry and surface abundance distribution of 36 Lyn]{Mixed poloidal-toroidal magnetic configuration and surface abundance distributions of the Bp star 36 Lyn\thanks{Based on observations obtained at the Telescope Bernard Lyot (USR5026) operated by the Observatoire Midi-Pyrénées, Université de Toulouse (Paul Sabatier), Centre National de la Recherche Scientifique of France.}}
\author[M.E. Oksala et al.]{M.~E. Oksala$^{1,2}$\thanks{E-mail:moksala@callutheran.edu},
J. Silvester$^{3}$, O. Kochukhov$^{3}$, C. Neiner$^{1}$, G.~A. Wade$^{4}$, 
\newauthor and the MiMeS Collaboration \\
$^{1}$LESIA, Observatoire de Paris, PSL Research University, CNRS, Sorbonne Universit\'es, UPMC Univ. Paris 06, Univ. Paris Diderot, \\
Sorbonne Paris Cit\'e, 5 place Jules Janssen, F-92195 Meudon, France \\
$^{2}$Department of Physics, California Lutheran University, 60 West Olsen Road \#3700, Thousand Oaks, CA, 91360, USA \\
$^{3}$Department of Physics and Astronomy, Uppsala University, Box 516, Uppsala 75120, Sweden \\
$^{4}$Department of Physics, Royal Military College of Canada, P.O. Box 17000, Station Forces, Kingston, Ontario K7K 7B4, Canada\\
}

\date{Accepted XXX. Received YYY; in original form ZZZ}

\pubyear{2017}

\begin{document}
\label{firstpage}
\pagerange{\pageref{firstpage}--\pageref{lastpage}}
\maketitle

\begin{abstract}

Previous studies of the chemically peculiar Bp star 36 Lyn revealed a moderately strong magnetic field, circumstellar material, and inhomogeneous surface abundance distributions of certain elements.  We present in this paper an analysis of 33 high-signal-to-noise ratio, high-resolution Stokes~$IV$  observations of 36 Lyn obtained with the Narval spectropolarimeter at the Bernard Lyot telescope at Pic du Midi Observatory.  From these data, we compute new measurements of the mean longitudinal magnetic field, $B_{\ell}$, using the multi-line least-squares deconvolution (LSD) technique.  A rotationally phased $B_{\ell}$ curve reveals a strong magnetic field, with indications for deviation from a pure dipole field.  We derive magnetic maps and chemical abundance distributions from the LSD profiles, produced using the Zeeman Doppler Imaging code {\sc InversLSD}.  Using a spherical harmonic expansion to characterize the magnetic field, we find that the harmonic energy is concentrated predominantly in the dipole mode ($\ell$ = 1), with significant contribution from both the poloidal and toroidal components.   This toroidal field component is predicted theoretically, but not typically observed for Ap/Bp stars.    Chemical abundance maps reveal a helium enhancement in a distinct region where the radial magnetic field is strong.  Silicon enhancements are located in two regions, also where the radial field is stronger.  Titanium and iron enhancements are slightly offset from the helium enhancements, and are located in areas where the radial field is weak, close to the magnetic equator.

\end{abstract}

\begin{keywords}
stars: magnetic fields - stars: chemically peculiar - stars: individual: 36 Lyn - techniques: spectroscopic - techniques: polarimetric
\end{keywords}



\section{Introduction}

Magnetic chemically peculiar late B- and A-type (Ap/Bp) stars  are characterized by large-scale, stable, often dipole-dominated magnetic fields and the presence of elemental surface inhomogeneities, or chemical ``spots''.  The appearance of such spot features is likely a consequence of the magnetic field modifying the typical mixing processes within the stellar outer envelope and atmosphere. The standard model that describes the variability of this class of stars is the oblique rotator model \citep{Babcock:1949aa,Stibbs:1950aa}, which describes a rotating star whose magnetic axis is at an angle, or obliquity, to its rotational axis.  By exploiting the rotation of these systems, Ap/Bp stars are typically studied using Zeeman Doppler imaging (ZDI) techniques to map the chemical surface and magnetic field structure of individual stars through a time-series of spectropolarimetric observations.  Recently, the number of Ap/Bp stars for which the magnetic field has been reconstructed using high-resolution spectropolarimetry has increased to $\sim$10 \citep{Koch14,Kochukhov15,Kochukhov:2017aa, Kochukhov10,Rusomarov15, Silvester14a,Silvester14b,Silvester15}. The magnetic geometry and the geometry of the chemical abundance inhomogeneities in Ap/Bp stars give key insight into the physical mechanisms governing the global magnetic field evolution in stellar interiors \citep{Braithwaite:2006aa,Duez:2010aa} and for testing theories of chemical diffusion \citep{Alecian15,Stift:2016aa,Alecian:2017aa}.

\begin{table*}
\centering
\caption{Log of the Narval spectropolarimetric observations, magnetic field measurements, and H$\alpha$ equivalent width (EW) measurements for 36 Lyn.  Column 1 gives the heliocentric Julian date (HJD) of mid-observation.  Column 2 lists the rotational phase according to the ephemeris determined in this work.  Column 3 provides the peak signal to noise ratio (SNR) per 1.8 \kms\, per spectral pixel in the reduced Stokes~$I$ spectrum.  Columns 4 and 5 report the SNR in the Stokes~$I$ and Stokes~$V$ LSD profiles for a velocity step of 3 \kms.  Columns 5 and 6 show the longitudinal magnetic field computed from LSD Stokes~$V$ ($B_\ell$) and $N$ ($N_\ell$) profiles, along with their associated formal errors ($\sigma_B$ and $\sigma_N$).  Column 7 lists the measured H$\alpha$ EW values and the associated 2$\sigma$ error. }
\begin{tabular}{lccccrrr}
\hline \hline
&&& && ~~LSD V~~~ & ~~LSD N~~~& \\
HJD & Phase  & Peak & LSD $I$ &  LSD $V$ & $B_\ell$ $\pm$ $\sigma_B~$ & $N_\ell$ $\pm$ $\sigma_N$~&   EW$_{\rm{H}\alpha} \pm \sigma_{\rm{EW}}$   \\ 
(2450000+)  &   & SNR  &SNR&SNR& ~~~~~(G)~~~~~ & ~~~~~(G)~~~~~ & ~~~~~(nm)~~~~~ \\ \hline
4902.4828 & 0.658  & 882  &  4830  &  23903 &  $-$526 $\pm$ 10       & +14 $\pm$ 9 & 1.330  $\pm$     0.012 \\
4903.5237 & 0.929  & 745  &  4263  &  21587 &  $-$217 $\pm$ 10       & $-$26 $\pm$ 10 & 1.540  $\pm$     0.012 \\
4904.5223 & 0.190  & 953  &  4912  &  27508 &  +738 $\pm$ 9    & $-$14 $\pm$ 8 & 1.317   $\pm$    0.012 \\
4905.4338 & 0.428  & 864  &  4155  &  22096 & +373 $\pm$ 10    & +11 $\pm$ 9 & 1.344    $\pm$   0.012 \\
4906.4630 & 0.696  & 717  &  4835  &  20723 & $-$600 $\pm$ 11          & +1 $\pm$ 11 & 1.290   $\pm$    0.012 \\
4907.5268 & 0.973  & 824  &  4447  &  24090 & +10 $\pm$ 9        & +2 $\pm$ 9 & 1.382    $\pm$   0.012 \\
4908.4887 & 0.224  & 835  &  4893  &  24159 &  +748 $\pm$ 10     & +24 $\pm$ 10 & 1.319    $\pm$   0.012 \\
4935.4306 & 0.250  & 774  &  4868  &  22070 &  +674 $\pm$ 11        & $-$12 $\pm$ 10 & 1.318    $\pm$   0.012\\
4935.4500 & 0.255  & 727  &  4825  &  20973 &  +667 $\pm$ 11      & +5 $\pm$ 11 & 1.307    $\pm$    0.012 \\
4945.3574 & 0.838  & 569  &  4408  &  16507 &  $-$594 $\pm$ 14        & +1 $\pm$ 14 & 1.331    $\pm$   0.012\\
4946.3642 & 0.101  & 820  &  4729  &  23904 & +630 $\pm$ 10    &  $-$8 $\pm$ 10 & 1.334   $\pm$    0.012 \\
4946.3839 & 0.106  & 843  &  4773  &  24269 &  +648 $\pm$ 10    & $-$6 $\pm$ 10 & 1.357    $\pm$    0.012 \\
5273.4543 & 0.395  & 575  &  4197  &  16156 &  +473 $\pm$ 13    & $-$1 $\pm$ 13 & 1.323    $\pm$    0.012 \\
5284.4219 & 0.255  & 558  &  4680  &  16023 &  +657 $\pm$ 14        & +18 $\pm$ 14 & 1.299    $\pm$    0.012\\
5292.4362 & 0.345  & 304  &  3984  &   8742  &  +553 $\pm$ 25           & $-$2 $\pm$ 25 & 1.309    $\pm$   0.012\\
5296.4216 & 0.384  & 785  &  4250  &   20722 & +494 $\pm$ 10       & $-$7 $\pm$ 10 & 1.322   $\pm$    0.012 \\
5298.4843 & 0.922  & 265  &  3735  &   6541  & $-$228 $\pm$ 33     & $-$56 $\pm$ 33 & 1.540   $\pm$    0.012\\
5304.4202 & 0.470  & 450  &  3952  &  12319 & +254 $\pm$ 17           & +1 $\pm$ 17 & 1.409   $\pm$    0.012\\
5312.4262 & 0.558  & 684  &  4515  &  18293 & $-$92 $\pm$ 12         &   +5 $\pm$ 11 & 1.317   $\pm$    0.012\\
5944.5814 & 0.403  & 769  &  4120  &   20082 & +449 $\pm$ 11       & +6 $\pm$ 10 & 1.322    $\pm$   0.012 \\
5950.5913 & 0.970  & 759  &  4313  &   21790 & +21 $\pm$ 10     & $-$20 $\pm$ 10 & 1.377    $\pm$   0.012\\
5951.5832 & 0.229  & 701  &  4721  &   19987 & +708 $\pm$ 12           & +1 $\pm$ 12 & 1.334      $\pm$  0.012\\
5952.6034 & 0.495  & 871  &  4119  &    21858 &+147 $\pm$ 10         &   $-$4 $\pm$ 9 & 1.380    $\pm$   0.012\\
5988.4787 & 0.850  & 785  &  4386  &    21873 & $-$514 $\pm$ 10       & $-$13 $\pm$ 10 & 1.341   $\pm$    0.012\\
5988.6739 & 0.901  & 532  &  4128  &    14621 & $-$311 $\pm$ 15     & $-$26 $\pm$ 15 & 1.546   $\pm$    0.012\\
5990.4925 & 0.375  & 691  &  4218  &    18470 & +496 $\pm$ 12           & $-$10 $\pm$ 11 & 1.353    $\pm$   0.012\\
5998.4363 & 0.447  & 477  &  3978  &    13246 & +318 $\pm$ 16         &   $-$28 $\pm$ 16 & 1.381    $\pm$   0.012\\
5999.4182 & 0.703  & 699  &  4765  &    19679 & $-$685 $\pm$ 12       & +11 $\pm$ 11 &  1.297   $\pm$    0.012\\
6001.5415 & 0.256  & 784  &  4619  &    22360 & +696 $\pm$ 10     & +6 $\pm$ 10 & 1.316    $\pm$   0.012 \\
6010.4042 & 0.567  & 545  &  4397  &    12655 & $-$136 $\pm$ 18           & $-$13 $\pm$ 17  & 1.327   $\pm$    0.012\\
6012.4619 & 0.104  & 831  &  4703  &    23678 & +656 $\pm$ 10         &   $-$8 $\pm$ 10  & 1.325   $\pm$    0.012\\
6304.6037 & 0.285  & 504  &  4352  &    14551 & +638 $\pm$ 16     & $-$46 $\pm$ 16 & 1.311    $\pm$   0.012\\
7368.6869 & 0.763  & 559  &  4137  &    15561 & $-$690 $\pm$ 15           & +2 $\pm$ 15 & 1.303    $\pm$   0.012\\
\hline \hline
\end{tabular}
\label{obs}
\end{table*}

36 Lyn (HD 79158, HR 3652) is a well-known magnetic Bp star, classified as B8IIImnp in the Bright Star Catalog \citep{Hoffleit:1995aa}.  Its chemical peculiarity was first noted by \citet{Edwards:1932aa}, and the star was identified as a member of the He-weak class of late B-type stars by \citet{Searle:1964aa}.  Subsequent studies of 36 Lyn's properties followed in optical spectroscopy \citep[e.g.,][]{Mihalas:1966aa,Sargent:1969aa,Cowley:1972aa,Molnar:1972aa,Ryabchikova:1996aa}, UV spectroscopy \citep{Sadakane:1984aa,Shore:1987aa,Shore:1990aa}, spectrophotometry \citep{Adelman:1983aa}, optical photometry \citep{Adelman:2000aa}, and radio \citep{Drake:1987aa,Linsky:1992aa}.  

The magnetic field of 36 Lyn was first detected in a study of He-weak stars by \citet[][hereafter B83]{Borra:1983aa}, which followed from a successful study detecting magnetic fields in the hotter He-strong stars \citep{Borra:1979aa}.   The two observations measuring circular polarization in the wings of the H$\beta$ line produced one clear longitudinal magnetic field detection at the 3.8$\sigma$ level ($B_\ell = 1315 \pm 340$ G).  \citet[][hereafter S90]{Shore:1990aa} followed up this discovery with a set of 22 H$\beta$ magnetic observations, confirming the detection by B83, measuring variability of $\pm900$~G in the $B_\ell$ values and suggesting that the field has a dominant dipolar component.  From examining the available IUE data, S90 also detected the presence of circumstellar material, due to the confinement of the weak stellar wind by the magnetic field.   

With the advent of newly developed instrumentation, \citet[][hereafter W06]{Wade:2006aa} obtained 11 circular polarization (Stokes~$V$) and 8 linear polarization (Stokes~$QU$) observations with the MuSiCoS spectropolarimeter (R$\sim$35000), as well as 3 additional H$\beta$ polarimetric observations.  There was no significant signal detected in the linear polarization observations, but the Stokes~$V$ signatures and their variability were clear.  The study found overabundant surface spots of Fe and Ti and a complex (i.e., not purely dipolar) magnetic field.  Using the same MuSiCoS spectra and IUE spectra, \citet{Smith:2006aa} confirmed the presence of circumstellar material, and determined the material was located in a disk structure, similar to structures reported previously to exist around hotter magnetic Bp stars, such as $\sigma$ Ori E \citep[e.g.,][]{Walborn:1974aa,Groote:1976aa}.

In this paper, we present an analysis of 33 new high-resolution circular polarization observations of 36 Lyn.  Section\,\ref{OBS} describes the obtained observations and their reduction.  The computation of the longitudinal magnetic field, including an analysis of the rotational period, is presented in Section\,\ref{LMF}.  In Section\,\ref{sect:inversions}, we describe the Doppler mapping technique and present the results of the magnetic field structure and chemical abundance distributions.  We discuss our results and make final conclusions in Section\,\ref{DISC}.

\section{Observations}\label{OBS}

We have obtained 33 high-resolution (R$\sim$65000) broadband (370-1040\,nm) circular polarization spectra of 36 Lyn using the Narval spectropolarimeter on the 1.9-m Bernard Lyot telescope (TBL) at the Pic du Midi Observatory in France.  Each observation consisted of four sub-exposures taken with different configurations of the polarimeter. The total exposure time for each observation was 4x360s, with $\sim$40s readout time between each exposure.  The observations were obtained over the period from March 2009 to January 2016.  The observation details are given in Table~\ref{obs}.    

The reduction of the data was performed with the Libre-ESpRIT package \citep{Donati:1997aa}, including bias, flat-field, and ThAr calibrations.  The reduction constructively combines the sub-exposures to yield the Stokes~$I$ (intensity) and Stokes~$V$ (circular polarization) spectrum.   The sub-exposures are also combined destructively to produce the null ($N$) spectrum, which diagnoses any spurious contribution to the polarization measurement.

The reduced intensity spectra were normalized order by order using spline functions in the task \textit{continuum} within IRAF\footnote{IRAF is distributed by the National Optical Astronomy Observatory, which is operated by the Association of Universities for Research in Astronomy (AURA) under a cooperative agreement with the National Science Foundation.}.  The resulting normalization functions were then applied to the corresponding orders in the Stokes~$V$ and $N$ spectra.

\section{Longitudinal magnetic field}
\label{LMF}

\subsection{$B_\ell$ measurements}

The least-squares deconvolution multi-line technique \citep[LSD;][]{Donati:1997aa} was applied to the normalized spectra to produce average Stokes~$I$, $N$, and Stokes~$V$ line profiles.   These profiles are obtained using an inversion technique which models the observations as a convolution of a mean profile and a line mask containing the wavelength, line depth, and Land\'e factor of all the lines in a given wavelength range.  The resulting mean LSD profiles (interpreted in this section as single line profiles with a characteristic depth, wavelength, and Land\'e factor) are significantly higher in signal-to-noise ratio (SNR) than individual lines (Table~\ref{obs}).   The LSD procedure propagates the formal error associated with each spectral pixel throughout the deconvolution process.  

\begin{figure*}
\centering
\includegraphics[width=5 in]{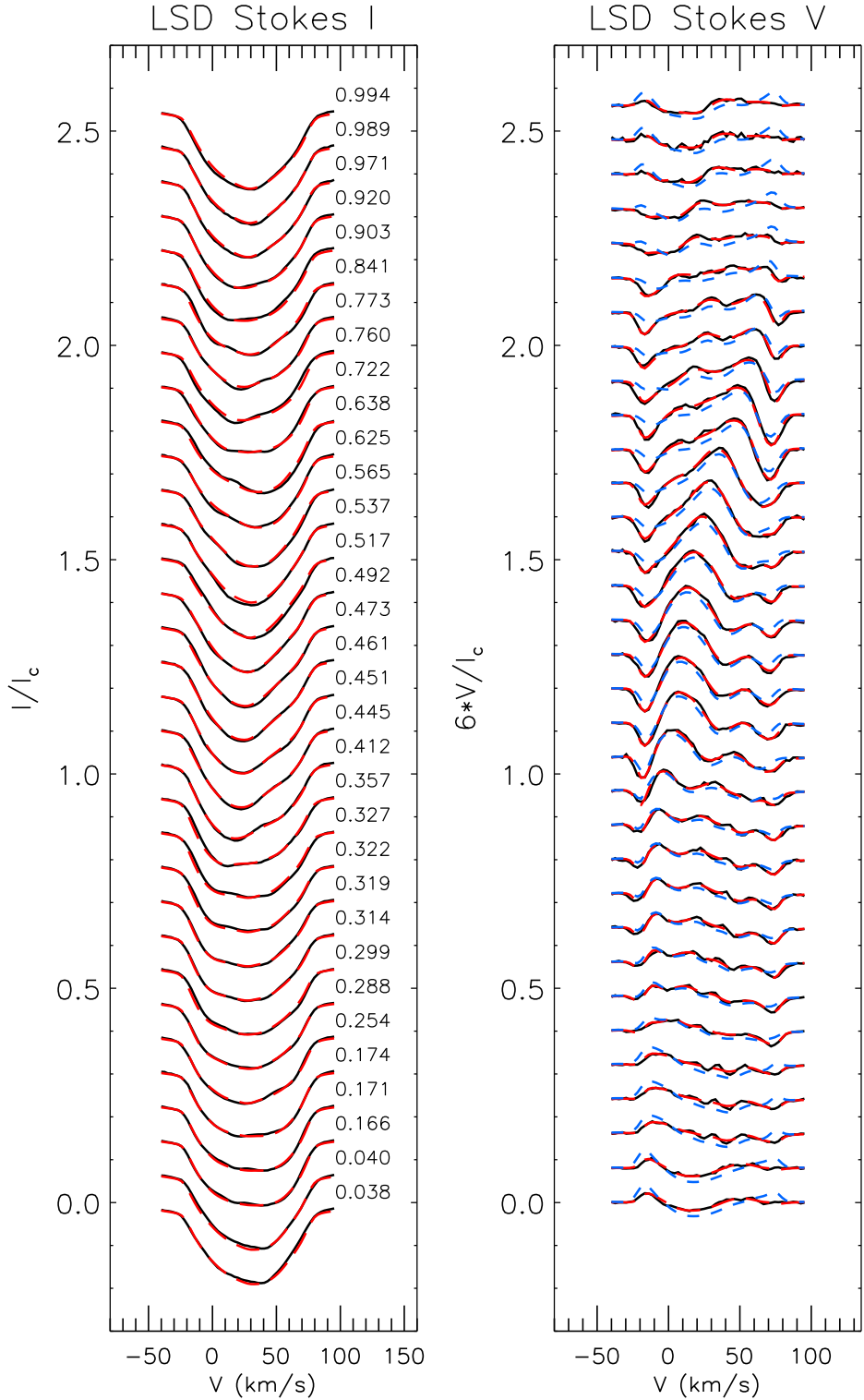}
\caption{Comparison between the observed Fe LSD Stokes $IV$ profiles for 36 Lyn and the fits obtained from the magnetic field inversion.  The LSD profiles plotted here were produced using iLSD \citep{Kochukhov:2010ab}, as described in Sect.~\ref{sect:inversions}.  The observations are given by the thick dark line.  The model fit including a toroidal field is given by the dashed red line and the model without a toroidal field is given by the dashed blue line. The number on the right of the Stokes $I$ profile indicates phase of rotation.}
\label{fieldfit}
\end{figure*}

For 36 Lyn, the line mask was developed starting from a VALD line list \citep{Piskunov:1995aa,Ryabchikova:2015aa} corresponding to \teff\ = 13000\,K and \lgg\ = 4.0, and assuming solar abundances.  The line list included all lines with predicted depths greater than 10\% of the continuum.  The line mask was subsequently cleaned of all hydrogen lines, lines blended with hydrogen, and regions of telluric contamination.  This ``cleaned'' mask was then applied to each individual spectrum, after which the mask was ``tweaked'' to adjust individual line depths to their proper values.  This tweaking allows, in particular, to account for non-solar abundances. This procedure is explained in detail by \citet{Grunhut:2017aa}.  All the tweaked masks were then averaged to produce a final mask, which was used to obtain the final LSD line profiles.  Approximately 1630 spectral lines were averaged to obtain the final profiles, dominated by Fe features, but containing a multitude of metal (Ti, Si, O, C, Ne, etc.) and helium lines.  The null profiles of all observations were flat, and the Stokes~$IV$ LSD profiles computed using solely Fe features ($\sim$950 Fe lines), but nearly identical to those obtained with all lines, are shown in Fig.~\ref{fieldfit}.

From the LSD-averaged Stokes~$I$ and $V$ line profiles, the longitudinal magnetic field is computed using the first moment of Stokes~$V$ about its center of gravity: 
\begin{equation}
B_{\ell} = -2.14 \times 10^{11} \frac{ \int v V(v) dv}{\lambda g c \int [I_{c}-I(v)]dv}~~~G  ,
\label{Bleq}
\end{equation}
\noindent where $I_c$ is the continuum value of the Stokes~$I$ profile, $g$ is the Land\'{e} factor (SNR-weighted mean value for all the included lines) and $\lambda$ is the mean wavelength in nm \citep{Donati:1997aa,Wade:2000ab}.  This value is similarly computed for the LSD-averaged $N$ profile.  The integration range that was used in this computation was $\pm~$60~\kms\ from the measured central point of the average intensity profile.   The computed values of $B_\ell$ for Stokes~$V$ and $N$ are listed in Table~\ref{obs}, along with their corresponding error bars.  Each Stokes~$V$ observation was found to be a definite detection (DD) based on the false alarm probability (FAP) discussed by \citet{Donati:1992aa,Donati:1997aa}. 

\subsection{Period analysis}\label{period}

With the addition of a new epoch of data, the rotational period of 36 Lyn was revisited to determine any possible changes, or more likely, any improvement in precision. As a first step, we analyzed the magnetic data from all possible epochs to achieve the longest possible time baseline.  The computed longitudinal magnetic field values ($B_\ell$) and their corresponding error bars ($\sigma_B$) listed in Table~\ref{obs}, combined with the historical measurements of B83, S90, and W06, were input to the period searching program FAMIAS \citep{Zima:2008aa}.  FAMIAS can be used to perform a simple period search within the data set using a Fourier 1D analysis.  A least-squares fit to the data, using a Levenberg-Marquardt algorithm, produces the formal statistical uncertainties.   For the magnetic observations of 36 Lyn, the analysis results in a period of 3.83484 $\pm$ 0.00003 days.  This period was confirmed using Period04, which performs a similar Fourier analysis combined with a least-squares fit \citep{Lenz:2005aa}.   

The period determined from all of the available $B_\ell$ measurements is consistent with the period determined by S90 (3.8345 $\pm$ 0.001\,d) from their $B_\ell$ data, as well as the period determined from the photometric variation by \citet[][3.834~d]{Adelman:2000aa}.  It is, however, inconsistent with both of the periods determined by W06 from a similar $B_\ell$ analysis (P = 3.83495 $\pm$ 0.00003\,d) and from an analysis of the H$\alpha$ equivalent width (EW) variation (P = 3.83475$\pm$0.00002\,d).  W06 find that the period determined from the magnetic data did not phase the H$\alpha$ variation coherently, and therefore they adopted the shorter period associated with the H$\alpha$ variation as the rotational period of the star.

\begin{figure}
\centering
\includegraphics[width=3.4 in]{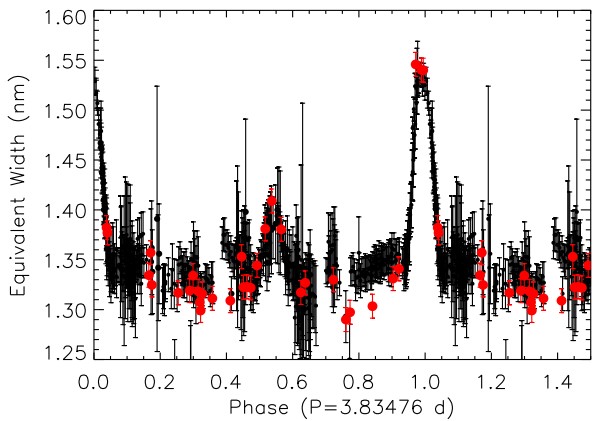}
\caption{H$\alpha$ equivalent width measurements for 36 Lyn.   The plot includes both data from this paper (larger red filled circles), as well as historical data reported by W06 (smaller black filled circles).  Measurements from this paper are listed in Table 1.The data are phased according to the ephemeris discussed in Section~\ref{period}.
  }
\label{Halpha}
\end{figure}

\begin{figure}
\centering
\includegraphics[width=3.4 in]{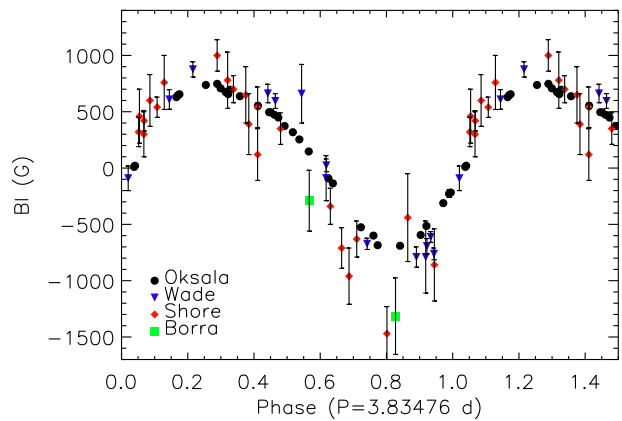}
\caption{Longitudinal magnetic field measurements for 36 Lyn.   Plotted are the measurements with corresponding error bars for the Narval dataset presented in this paper (black filled circles), as well as the data reported by W06 (blue triangles), S90 (red diamonds) and B83 (green squares).   The data are phased according to the ephemeris discussed in Section~\ref{period}.  }
\label{Blall}
\end{figure}

\begin{figure}
\centering
\includegraphics[width=3.4 in]{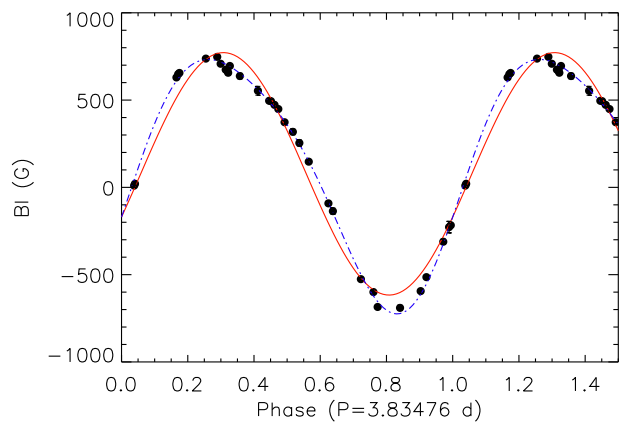}
\caption{Longitudinal magnetic field measurements (black dots) for 36 Lyn determined from the new dataset presented in this paper.  A second order least-squares harmonic fit (dot-dashed blue curve) is clearly in better agreement with the observations than a first order fit (solid red curve), indicating that the magnetic field may contain higher order (non-dipolar) components. The data are phased according to the ephemeris discussed in Section~\ref{period}.
  }
\label{Blfit}
\end{figure}

\begin{figure*}
\begin{center}
    \includegraphics[width=0.74\textwidth, angle=-90]{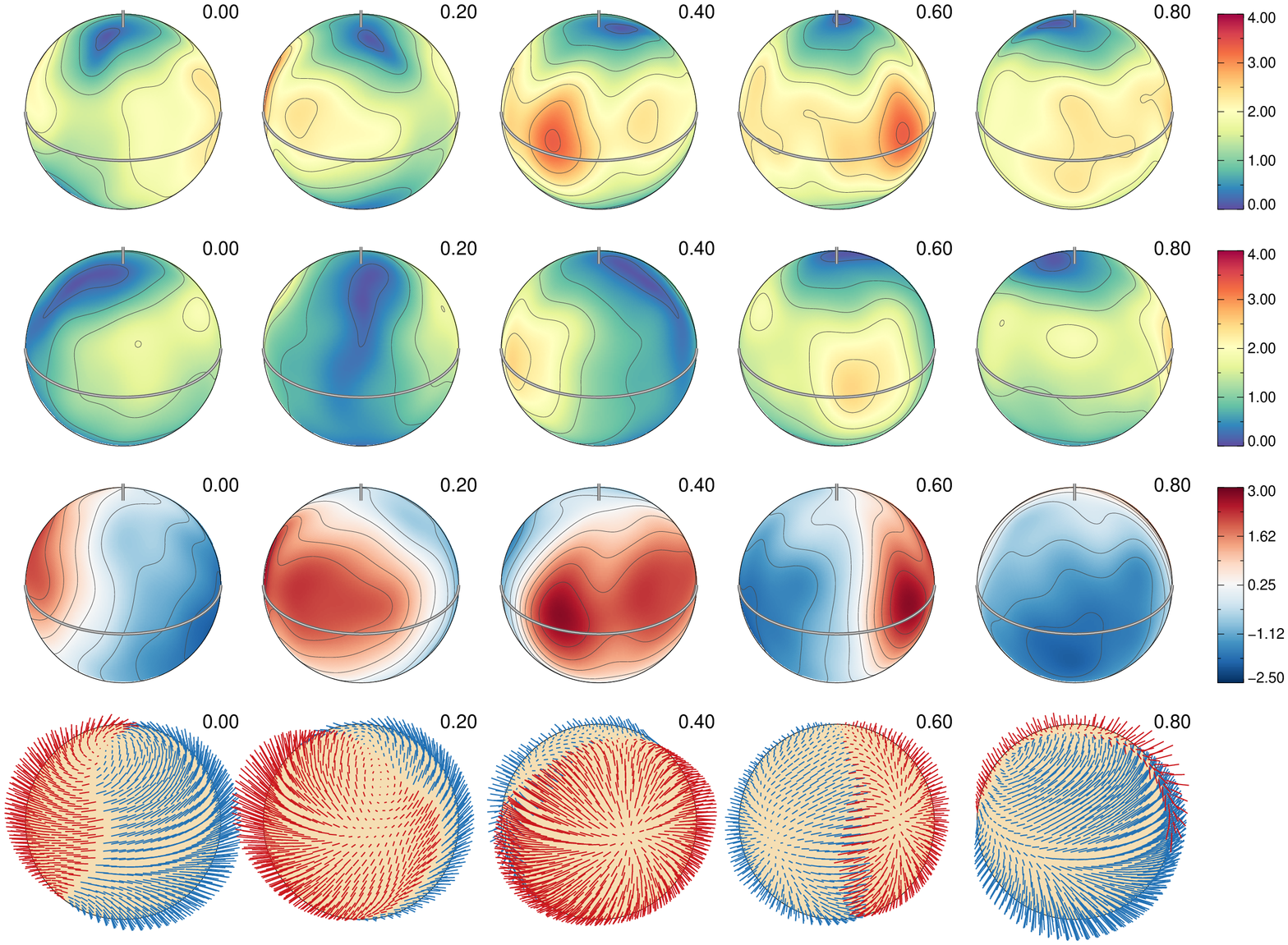}
    \includegraphics[width=1.44 in, angle=-90]{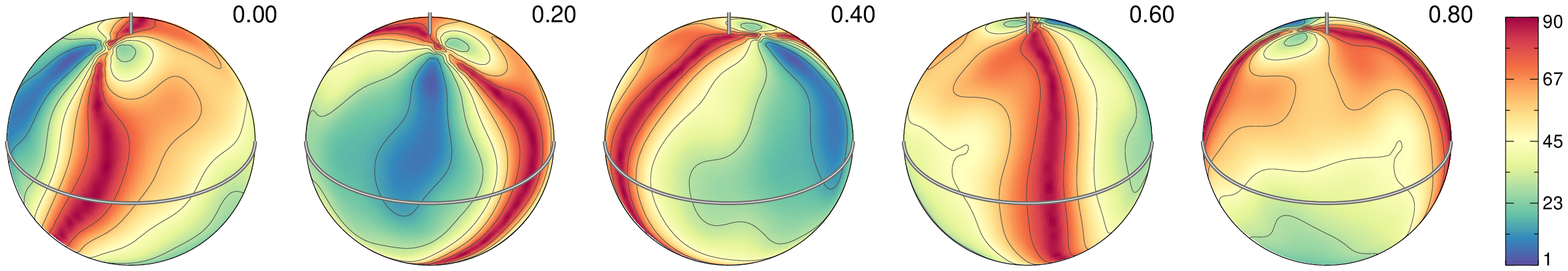}
  \caption{Surface magnetic field distribution of 36 Lyn derived from LSD Stokes $IV$ profiles with {\sc InversLSD}. The spherical plots (at inclination angle of $ i = 60\degr$) show distributions of: the field modulus (top), the horizontal field (second from top), the radial field (third from top),  the field orientation (second from bottom), and the angle between the local field vector and the surface normal (bottom).  Each column corresponds to a different phase of rotation (0.0, 0.2, 0.4, 0.6, and 0.8) and the color bars on the right side indicate the magnetic field strength in kG (top three rows) or the angle in degrees (bottom).}
\label{maps-field}
\end{center}
\end{figure*}

With the new high-resolution Stokes~$I$ spectra in this study, we investigated the period obtained from the variation of the EW of H$\alpha$, similar to the analysis strategy of W06.  Adding to the set of H$\alpha$ data presented by W06 (see their Table 1 for more information), we analyzed the entire set of EW measurements, including the 33 new data points from our spectra, whose values and error bars were obtained using the same procedure as W06 (listed in Table~\ref{obs}).  The advantage to using H$\alpha$ line variability as a period diagnostic is that the core of the line changes depth quite sharply and consistently throughout the several decades of observations, as a result of the eclipsing of the stellar disk by circumstellar material within the magnetosphere.  The period of rotation of the star can therefore be measured with high precision, as any shift in period would cause data to scatter from the sharp peak.  We find that a period of 3.83476 $\pm$ 0.00004 provides the best fit for this multi-epoch data, as shown in Fig.~\ref{Halpha}.  Our new measurements fit well with the historic data, particularly in the location of the sharp peak in EW.  The error bars were determined from the eventual obvious shift of data out of phase, given the long time baseline.  This period is consistent with the value found by W06 from their H$\alpha$ EW analysis.

Least-squares harmonic fits to the magnetic data were obtained using as fixed frequencies both the rotational period determined from the H$\alpha$ EW measurements and the period obtained from our initial analysis using FAMIAS.  The fit using the H$\alpha$ EW-determined period yields a lower $\chi_{\rm{red}}^2$ value (3.75 vs. 4.75) than for the fit using the period determined using solely the magnetic data.  The EW-determined period is therefore consistent, phasing well both the EW variations and the magnetic measurements over a time span of $\sim$38 years.  The variability of the magnetic field is more subtle than the sharp peak of EW for H$\alpha$, possibly contributing to the erroneous results of the FAMIAS and PERIOD04 programs.  In addition, the magnetic field measurements have been obtained with not only a variety of measurement techniques, but also different details of the analysis methods (i.e., a different line mask for our data vs. that used by W06).   Inconsistency could lead to ambiguity in the rotational period, and the more homogeneous sample of H$\alpha$ measurements produces a dependable and more accurate determination. 

For the remainder of our analysis, we therefore adopt a rotational ephemeris of:
\begin{equation}
 JD = (2443000.451 \pm 0.03) + (3.83476 \pm 0.00004)~\rm{d}, 
\end{equation} 
where the HJD$_0$ value was determined from the H$\alpha$ variability by W06.

\subsection{$B_\ell$ curve}\label{BlLC}

Figure~\ref{Blall} shows the entire set of historic longitudinal magnetic field data plotted according to the above determined ephemeris.  The plot shows good agreement between current and historical measurements, although there is ambiguity as to the minimum value.  The error bars of the historical data, however, are larger by factors ranging from 5 to 20.  There may be additional disagreement between the measurements which consider metal and helium lines (this work and W06) versus those obtained from the wings of H$\beta$ (S90 and B83).   Larger longitudinal values obtained from hydrogen lines have been documented in a number of studies \citep[e.g.,][]{Bohlender:1987aa, Yakunin:2015aa}.

Considering only the new dataset of $B_\ell$ measurements and the determined ephemeris, a first order least-squares harmonic fit was compared with the data, and is plotted as the solid red curve in Fig.~\ref{Blfit}.  This simple sinusoid does not well fit the data and results in a poor fit with $\chi_{\rm{red}}^2 = 42.7$.  The data is better fit by a second order fit, shown in the same figure as a blue dashed-dot curve, which gives $\chi_{\rm{red}}^2 = 4.5$.  Comparison of these fits implies that the magnetic field is not well represented by a pure dipole structure, and that higher order (i.e. quadupole, octopole, etc.) components may be present.  The longitudinal field measurements, however, while useful to determine the presence of a magnetic field, do not allow us to determine the details of the magnetic field configuration.  In the next section, we describe the procedure to extract this information from a ZDI analysis of the Stokes~$I$ and $V$ line profiles.

\begin{figure}
\begin{center}
    \includegraphics[width=0.50\textwidth]{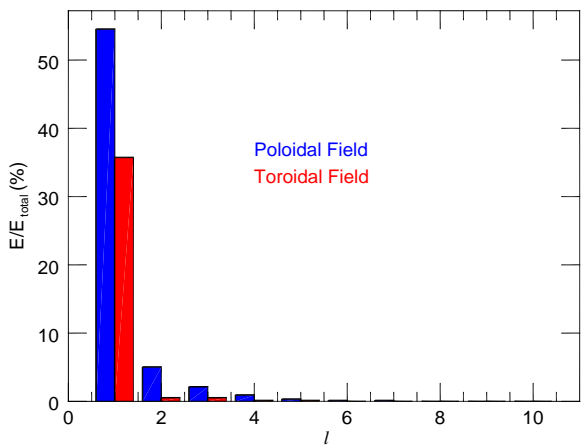}
   \caption{Distribution of the magnetic energy between different spherical harmonic modes in the magnetic field map of 36 Lyn.  Bars of different colors represent the poloidal (blue/dark) and toroidal (red/light) magnetic components.}
\label{power-plot}
\end{center}
\end{figure}

\section{Magnetic and chemical inversions}
\label{sect:inversions}

\subsection{Inversion technique}


\begin{figure*}
\begin{center}
 \vspace{5 mm}
 \includegraphics[width=0.62\textwidth, angle=-90]{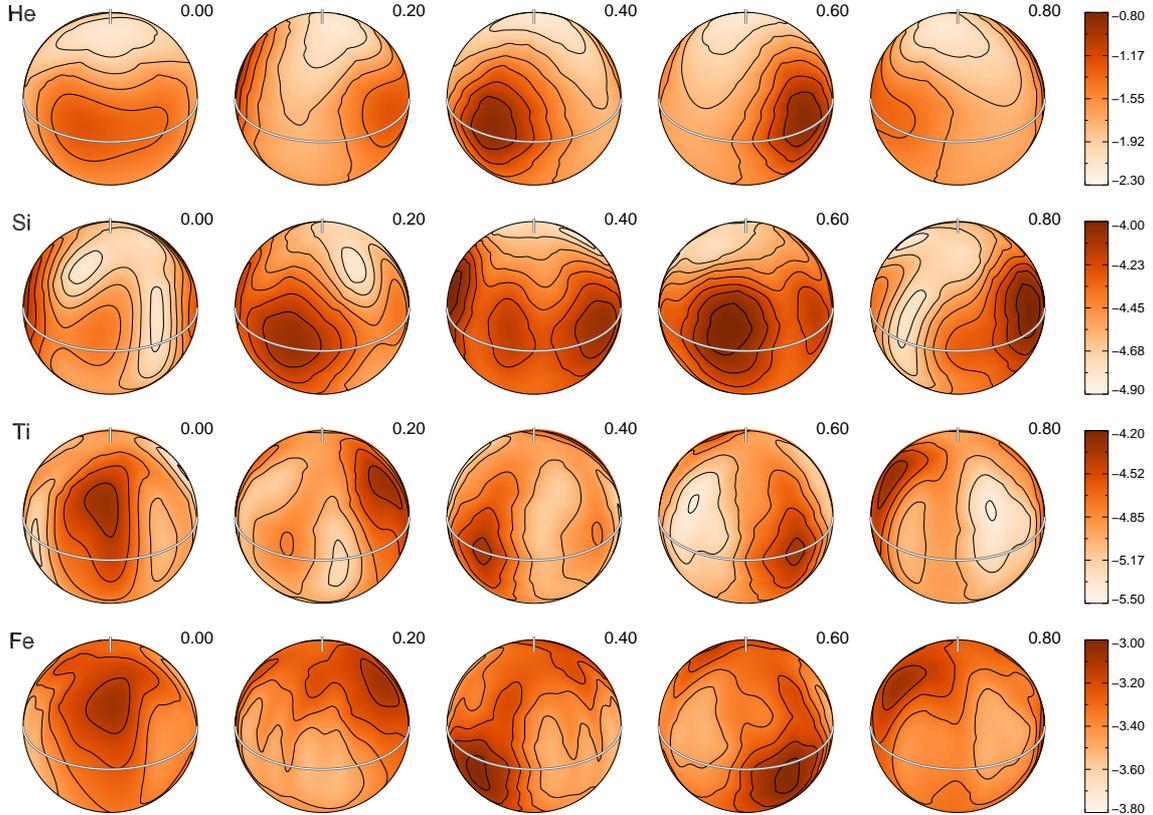}
    \caption{Chemical abundance distributions derived for He, Si, Ti (from individual lines), and Fe (from LSD profiles).  Each column corresponds to a different rotational phase (0.0, 0.2, 0.4, 0.6, and 0.8). The thick double white line shows the location of the stellar equator. The visible rotational pole is indicated by the short vertical white line. The color bars on the right side indicate chemical abundance in $\log N_{\rm el}/N_{\rm tot}$ units. }
\label{maps-abn}
\end{center}
\end{figure*}


\begin{figure}
\begin{center}
\vspace{12mm}
 \includegraphics[width=2.1 in]{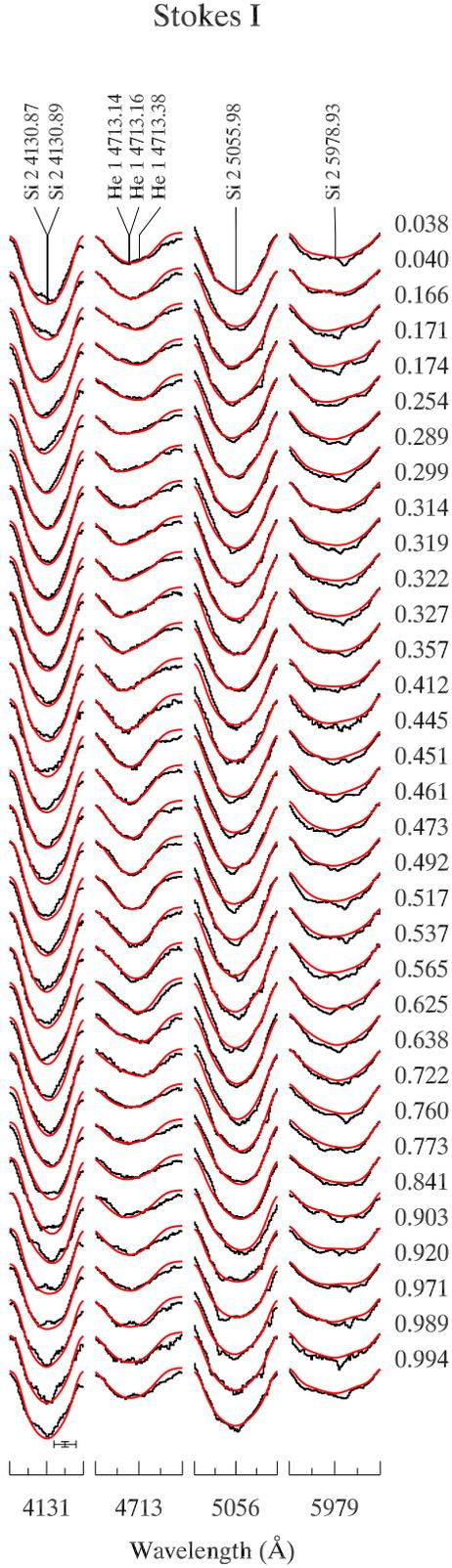}
 \vspace{15mm}
   \caption{Comparison between the observed (black line) and synthetic (red line) Stokes $I$ spectra of He and Si used for abundance mapping of these elements. The bars in the lower left corners indicate the vertical and horizontal scales (1 \% of the Stokes $I$ continuum intensity and 0.5 \AA). The rotational phase is indicated to the right.}
\label{He-Si-fit}
\end{center}
\end{figure}

\begin{figure}
\begin{center}
\vspace{12mm}
 \includegraphics[width=1.9 in]{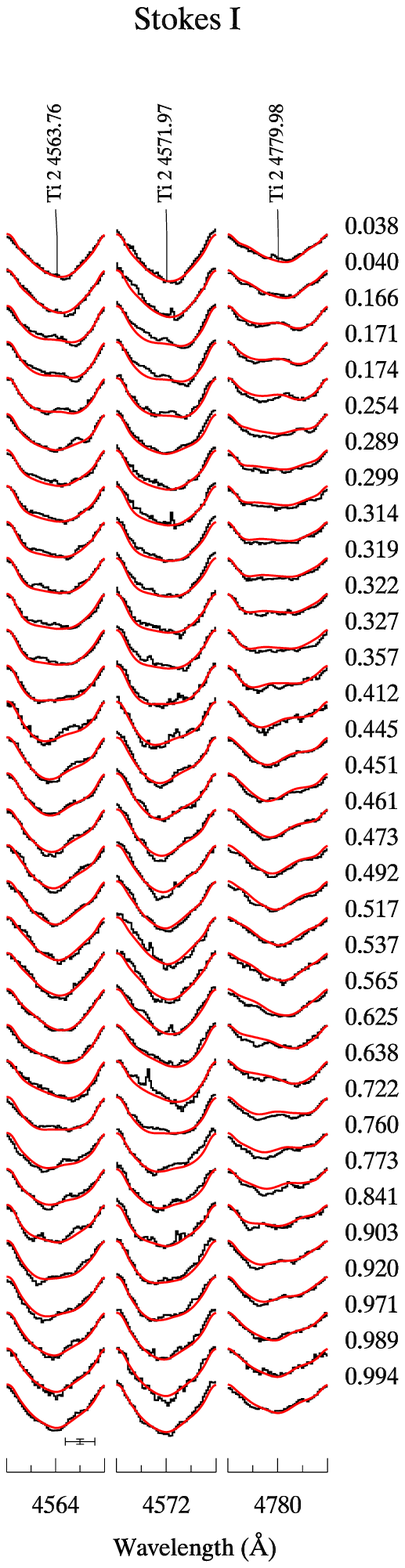}
 \vspace{15mm}
  \caption{Same as Fig. \ref{He-Si-fit}, but for Ti.}
\label{Ti-fit}
\end{center}
\end{figure}

To reconstruct the magnetic field of 36 Lyn, we performed inversions of Stokes $IV$ using {\sc InversLSD}, a version of the INVERS code which utilizes LSD averaged line profiles.  The details of this code are described by \citet{Koch14} and \citet{Rosen15}.  LSD profiles were re-derived using iLSD \citep{Kochukhov:2010ab}, a code which reconstructs multiple mean profiles from a single spectrum and uses regularization to reduce numerical noise within the LSD profile. The utilization of iLSD allows for consistency and the correct formatting for inclusion into the inversion input files. To extract the Fe iLSD profiles, we used a line mask containing 1480 Fe lines.  As discussed by \citet{Rosen15}, {\sc InversLSD} uses interpolated local profiles which have been pre-computed, rather than performing the polarized radiative transfer calculations in real time, allowing for a quick solution convergence. Within the inversion code the magnetic field topology is parametrized in terms of a general spherical harmonic expansion (up to the angular degree $\ell=10$ in this study), which utilizes real harmonic functions, again described by \citet{Koch14}. To facilitate convergence to a global $\chi_{\rm{red}}^2$ minimum a regularization function is used, in the case of spherical harmonics this function limits the solution to the simplest harmonic distribution that provides a satisfactory fit to the data.  A map of one additional scalar parameter, in this case Fe abundance, is reconstructed simultaneously with the magnetic field with the aid of Tikhonov regularization \citep[e.g., ][]{Piskunov:2002aa}.  The spherical harmonic coefficients in the final model (truncated at $\ell = 5$) are listed in Table ~\ref{shc}.

Mapping the stellar surface requires accurate values of the projected rotational velocity (\vs) and the inclination angle ($i$, between the rotational axis and the line of sight) as input.  To determine these properties, we started with those reported by \citet{Wade:2006aa} and investigated, within the uncertainty range reported, which values gave the smallest deviation between the model spectra and the observations.  We found the following parameters: $i=60\degr \pm 10\degr$ and \vs\,=\,$49.0 \pm 1.0 $ \kms.  These values are consistent with those found by W06 ($i \geq 56\degr$ and \vs\,=\,$48 \pm 3 $ \kms).

For the final inversions, the level of regularization implemented was chosen by a stepwise approach.  Initially, the regularization was set to a high value, but with each inversion, or each successive step, the restriction or value of regularization decreased \citep{Kochukhov:2017ab}.  The final regularization level was chosen at the point where the fit between the observations and the model was no longer significantly improved by further decrease of the restriction. This approach was used both for the magnetic and abundance maps. 

With the final magnetic field geometry (Fig.~\ref{maps-field}) determined from the LSD inversions (Fig.~\ref{fieldfit}), we then derived surface chemical abundance maps modeling Stokes~$I$ profiles of individual spectral lines with the {\sc invers10} code \citep[e.g.,][see Fig.~\ref{maps-abn}]{Piskunov:2002aa,Kochukhov:2002aa}.  Because of the complexity of dealing with blends, we focused on unblended lines, resulting in a very small number of elements for which we could derive chemical maps. The final abundances mapped were He, Si, Ti (derived from individual line profiles), and Fe (derived from LSD profile inversions). The method to derive the abundance maps took the magnetic field previously derived from the average line inversions and included its influence on the line formation in the abundance mapping, with the {\sc invers10} code only adjusting the abundances.

\begin{table}
\begin{center}
\caption{Table of spherical harmonic coefficients for the magnetic field map of 36 Lyn truncated at $l=5$.  The expansion is formulated with real spherical harmonic functions, as described by \citet{Koch14}.}
\begin{tabular}{lcccc}
\hline
\hline
  l  &  m   &  $\alpha$   &   $\beta$  &   $\gamma$ \\
\hline
   1 & -1 &  5.26E+00 & 1.40E+00 & 1.03E+00 \\
   1 &  0  & 8.93E-01  & 1.01E+00 & 4.96E+00 \\
   1 &  1  & 2.67E+00  & 2.07E+00 & -1.67E+00 \\
   2 & -2  & 2.62E-01  & 1.49E+00 & 2.32E-01 \\
   2 & -1  & 4.94E-01 & -8.53E-01 & -2.28E-01 \\
   2 &  0 & -6.76E-01 & -1.24E-02  & 4.24E-01 \\
   2 &  1 & -2.27E-01  & 5.63E-02 & -1.33E-01 \\
   2 &  2  & 5.94E-01  & 4.47E-01 & -2.83E-01 \\
   3 & -3  & 2.41E-01 & 1.57E-01 & 2.25E-01 \\
   3 & -2  & 1.68E-01  & 2.16E-01 & -1.79E-01 \\
   3 & -1 & -3.94E-01 & -4.55E-01 & -2.73E-01 \\
   3 &  0 & -2.66E-01  & 7.10E-02 & -1.87E-01 \\
   3 &  1 & -1.28E-01 & -3.89E-01 & -3.10E-01 \\
   3 &  2 & -1.36E-01 & -1.12E-01 & 3.61E-01 \\
   3 &  3 & 8.65E-01  & 1.59E-01 & -6.85E-02 \\ 
   4 & -4 & 2.21E-01 & 1.69E-01 & 9.87E-02 \\ 
   4 & -3  & 6.70E-02 & 3.26E-02 & -8.28E-02 \\
   4 & -2  & 8.14E-02 & 5.54E-02 & -3.71E-02 \\
   4 & -1 & -2.71E-01 &-9.34E-02 & -1.80E-01 \\
   4 &  0  & 7.09E-02  & 1.03E-01 & -1.29E-01 \\
   4 &  1  & 1.23E-01 & -2.55E-01 & -1.73E-01 \\
   4 &  2 & -2.38E-01 & -1.22E-01 &  3.98E-02 \\
   4 &  3  & 2.64E-01 & -1.30E-01 & -3.33E-02 \\
   4 &  4  & 4.01E-01  & 3.36E-01 & -5.36E-02 \\
   5 & -5 & 1.01E-01  & 2.46E-02  & 6.41E-02 \\
   5 & -4  & 1.07E-02 & -3.22E-03 & -5.65E-02 \\
   5 & -3  & 1.09E-02 & 7.44E-02  & 1.10E-03 \\
   5 & -2  & 1.65E-02 & -3.65E-02  & 6.63E-02 \\
   5 & -1 & -8.56E-02 & -1.95E-02 & -5.91E-02 \\
   5 &  0 & 6.78E-02 & 1.88E-02 & -1.55E-01 \\
   5 &  1 & 8.98E-02 & -6.24E-02 & -1.42E-02 \\
   5 &  2 & -1.35E-01 & -2.09E-02 & -1.83E-02 \\
   5 &  3  & 2.74E-02 & -6.92E-02 & -8.26E-02 \\
   5 &  4  & 6.31E-02 & 5.68E-02 & 9.11E-03 \\ 
   5 &  5  & 3.41E-01 & 1.40E-01 & -4.44E-04 \\
 \hline 
   \label{shc}
\end{tabular}
\end{center}
\end{table}

\subsection{Magnetic field structure}
\label{sect:mag}

The final magnetic field map shown in Fig.~\ref{maps-field} reveals a relatively simple structure.    Overall, the magnetic field is dipolar in structure with prominent radial field regions located near the rotational equator at phases 0.4 and 0.6.   A graph of the harmonic energy levels in shown in Fig.~\ref{power-plot}, with these values in each harmonic reported in Table~\ref{energies}.  90\% of the magnetic field energy is in the first harmonic ($\ell=1$), confirming that the field geometry is primarily dipolar.  However, the dipolar ($\ell=1$) harmonic has a large toroidal component (36\%); this is a much larger contribution than has been observed to date in other Ap/Bp stars that have been mapped with a similar method \citep[e.g.,][]{Koch14,Rusomarov15,Rusomarov16,Silvester15}.  This toroidal component can also be identified in the vector field map in Fig.~\ref{maps-field},  where the magnetic vectors mark out a ring-like pattern around the star,  somewhat parallel to the equator.  This toroidal component is necessary to properly fit the observed Stokes~$V$ line profiles, and a significantly poorer fit is obtained without its inclusion. This is illustrated in Fig.~\ref{fieldfit} for the Fe LSD profiles, where the model without a toroidal component exhibits a poorer fit to the data.  Likewise, the profiles cannot be fit by limiting the inversions to a purely dipolar ($\ell=1$) or dipolar plus quadrupolar ($\ell=2$) structure. 


\begin{table}
\begin{center}
\caption{Distribution of poloidal ($E_{\rm pol}$), toroidal ($E_{\rm tor}$), and total magnetic field energy over different harmonic modes for the best-fitting magnetic field map of 36 Lyn.  }
\begin{tabular}{lccc}
\hline
\hline
$\ell$ & $E_{\rm pol}$ & $E_{\rm tor}$  & $E_{\rm tot}$\\
 & (\%) & (\%) & (\%) \\
\hline
  1   & 54.5 & 35.7  & 90.2 \\
  2  &  5.0  &  0.5  & 5.5 \\
  3   & 2.0   & 0.5 &2.5   \\
  4   &  0.9 &  0.1   & 1.0 \\
  5   & 0.3   & 0.1     & 0.4  \\
  6    & 0.1   & 0.0     & 0.1 \\
  7    & 0.1   &  0.0   & 0.1 \\
  8    & 0.0  & 0.0     & 0.0 \\
  9    & 0.0   & 0.0    & 0.0 \\
 10    & 0.1   & 0.0     & 0.1 \\
\hline
 Total &  63.4 &  36.6  & 100  \\
\hline
\label{energies}
\end{tabular}
\end{center}
\end{table} 

\subsection{Chemical abundance distributions}
\label{sect:chemabn}

\subsubsection{Helium and silicon}

The helium abundance distribution obtained from the Stokes $I$ profile of the He~{\sc i} $\lambda$ 4713 line (Fig.~\ref{maps-abn}) varies from -0.80 to -2.30 dex in $\log N_{\rm el}/N_{\rm tot}$ units (the solar value is -1.11 dex), with a large enhancement located near the rotational equator at a phase of 0.40. This enhancement appears to overlap with the strongest positive radial field feature in the magnetic map (Fig.~\ref{maps-field}). The fit of the model profiles to the observations is shown in Fig.~\ref{He-Si-fit}. 

The silicon abundance distribution inferred from the Stokes $I$ line profile fit to the Si~{\sc ii} $\lambda$ 4130, 5055, and 5978 lines varies from -4.00 to -4.90 dex (solar = -4.53 dex). There are 3 large areas of slight enhancement (compared to the solar abundance) around the stellar equator. These enhancements seem to be located in areas where the radial magnetic map is strong (Fig. \ref{maps-field}).  Again, the fits of the model profiles to the observations are shown in Fig.~\ref{He-Si-fit}.

\subsubsection{Titanium and iron}

The titanium abundance distribution was determined using Stokes $I$ profiles of Ti~{\sc ii} $\lambda$ 4563, 4571, and 4779 lines.  The abundance ranges from -4.20 to -5.50 dex (solar = -7.09 dex).  The Ti map (Fig.~\ref{maps-abn}) shows slightly elongated structures of both enhancement and depletion.  The areas of the largest enhancement are located close to where the radial field is weak (the magnetic equator). The depleted regions appear to center on regions where the radial field is stronger. The fits of the model profiles to the observations are shown in Fig.~\ref{Ti-fit}. 

In the case of iron, the element distribution derived from the LSD model fit (Fig.~\ref{maps-abn}; ranging from -3.00 to -3.80 dex, solar = -4.54 dex) is similar to what is observed for titanium. Large enhancement areas are found close to the magnetic equator and also located near weak regions of the radial field.  The fit between the observations and the model profiles is shown in Fig.~\ref{fieldfit}.

\section{Discussion and conclusions}\label{DISC}

Current theoretical diffusions models by \citet{Alecian15} predict that abundance enhancements should be located in regions where the surface field is primarily horizontal, or in regions where the radial magnetic field is weak.  The  derived abundance map of He indicates enhanced abundance in a distinct region where the radial magnetic field is strong, which is incompatible with the theoretical predictions, although their models do not address He behavior specifically.  A similar effect is seen for the Si abundance, but with enhancements throughout the entire region of strong radial field, also in contrast with predicted diffusion behavior.   For the elements titanium and iron, the enhancements are slightly offset from the helium enhancements, located in areas where the radial field is weak, close to the magnetic equator, in qualitative agreement with the theoretical modeling by \citet{Alecian15}.

The magnetic field maps derived for 36 Lyn indicate that this star has a dipole-dominated field geometry, with a simpler structure compared to what is reported for other Ap/Bp stars mapped with four Stokes parameter data in a similar temperature range (e.g, HD 32633 and $\alpha^2$~CVn).  However, 36 Lyn does have a large toroidal component in the $\ell=1$ mode, something which has not typically been seen in such stars. Without the inclusion of Stokes~$QU$ spectra, however,  the derived field maps  are unable to reveal any small-scale structure that may also be present \citep{Kochukhov:2002aa}.

It was suggested by \citet{Silvester15} and \citet{Rusomarov16} that there is a slight trend for more complex fields to be found in hotter Ap/Bp stars and simpler fields to be found in cooler Ap stars (such as HD\,24712 and HD\,75049). The temperature of 36 Lyn places it in the same group as Bp stars such as  $\alpha^2$~CVn \citep{Kochukhov10,Silvester14a,Silvester14b},  CU Virginis \citep{Koch14},  HD\,32633 \citep{Silvester15}, and HD 125248 \citep{Rusomarov16}. Each of these stars were mapped using all four Stokes parameters, $IQUV$, with the exception of CU Virginis, which was studied with Stokes $IV$ observations.  For comparison, the magnetic field we derive for 36 Lyn is simpler than the configuration found for CU Vir, which has $\sim$20\% of its energy in the $\ell=2$ mode, compared to $\sim$5\% for 36 Lyn.  The complexity of the magnetic field of 36 Lyn lies somewhere between these aforementioned stars and the simpler geometry seen in cooler stars, e.g., the roAp star HD\,24712 \citep[DO Eri;][]{Rusomarov15} and HD\,75049 \citep{Kochukhov15}, with the notable exception that the large toroidal component is not typically detected in such cooler stars.

The detection of a significant toroidal component to the magnetic field structure provides observational validation of previous theoretical works that indicate purely poloidal and purely toroidal fields are unstable on their own \citep{Markey:1973aa,Braithwaite:2007aa}.  Simulations by \citet{Braithwaite:2009aa} and \citet{Duez:2010ab} \citep[incorporating the analytical theory developed by][]{Duez:2010aa} find that the ratio of poloidal magnetic energy to the total magnetic energy, $E_p/E$, is required to be between 0.04 and 0.8 to ensure that the field remains stable.  $E_p/E=0.634$ for the magnetic field of 36 Lyn, satisfying the theoretical mixed field requirement.  The presence of a \textit{surface} toroidal field component, however, is not currently predicted by these studies, and thus its observation is surprising.  

There are stars, such as HD 37776 \citep{Kochukhov:2011aa} and $\tau$~Sco \citep{Koch16}, which have very complex fields, and also show significant toroidal field components.  However, the contribution from harmonics larger than $\ell=1$ is significant in these stars. This makes 36 Lyn currently a unique case, combining a very simple geometry with a large toroidal component in $\ell=1$.  The extension of spectropolarimetric observations to Stokes~$QU$ parameters could potentially help confirm the authenticity of this large toroidal component.  However, the observations reported by W06 found no signal in linear polarization, indicating that for this star a much higher SNR in linear polarization is required for useful Stokes~$QU$ data. 

The question for future studies remains, why are we able to detect such a significant toroidal component for 36 Lyn, but not for similar stars, with similar magnetic topologies?  Work on this problem should consider whether all similar field topologies have significant toroidal components, and if so, in which particular situations are we are able to detect their observational signatures.

\section*{Acknowledgements}

MEO would like to thank Bram Buysschaert for illuminating discussions on period analysis.  JS would like to thank Lisa Ros{\'e}n for her advice. OK is a Royal Swedish Academy of Sciences Research Fellow supported by grants from the Knut and Alice Wallenberg Foundation, the Swedish Research Council and Goran Gustafsson Foundation.  GAW acknowledges Discovery Grant support from the Natural Sciences and Engineering Research Council (NSERC) of Canada. This reasearch has made use of the VALD database, operated at Uppsala University, the Institute of Astronomy RAS in Moscow, and the University of Vienna.  This research has also made use of the SIMBAD database, operated at CDS, Strasbourg, France and NASA's Astrophysics Data System Bibliographic Services.




\bibliographystyle{mnras}
\bibliography{36lyn} 

\begin{thebibliography}{}
\makeatletter
\relax
\def\mn@urlcharsother{\let\do\@makeother \do\$\do\&\do\#\do\^\do\_\do\%\do\~}
\def\mn@doi{\begingroup\mn@urlcharsother \@ifnextchar [ {\mn@doi@}
  {\mn@doi@[]}}
\def\mn@doi@[#1]#2{\def\@tempa{#1}\ifx\@tempa\@empty \href
  {http://dx.doi.org/#2} {doi:#2}\else \href {http://dx.doi.org/#2} {#1}\fi
  \endgroup}
\def\mn@eprint#1#2{\mn@eprint@#1:#2::\@nil}
\def\mn@eprint@arXiv#1{\href {http://arxiv.org/abs/#1} {{\tt arXiv:#1}}}
\def\mn@eprint@dblp#1{\href {http://dblp.uni-trier.de/rec/bibtex/#1.xml}
  {dblp:#1}}
\def\mn@eprint@#1:#2:#3:#4\@nil{\def\@tempa {#1}\def\@tempb {#2}\def\@tempc
  {#3}\ifx \@tempc \@empty \let \@tempc \@tempb \let \@tempb \@tempa \fi \ifx
  \@tempb \@empty \def\@tempb {arXiv}\fi \@ifundefined
  {mn@eprint@\@tempb}{\@tempb:\@tempc}{\expandafter \expandafter \csname
  mn@eprint@\@tempb\endcsname \expandafter{\@tempc}}}

\bibitem[\protect\citeauthoryear{{Adelman}}{{Adelman}}{2000}]{Adelman:2000aa}
{Adelman} S.~J.,  2000, \aap, \href
  {http://adsabs.harvard.edu/abs/2000A%26A...357..548A} {357, 548}

\bibitem[\protect\citeauthoryear{{Adelman} \& {Pyper}}{{Adelman} \&
  {Pyper}}{1983}]{Adelman:1983aa}
{Adelman} S.~J.,  {Pyper} D.~M.,  1983, \aap, \href
  {http://adsabs.harvard.edu/abs/1983A%26A...118..313A} {118, 313}

\bibitem[\protect\citeauthoryear{{Alecian}}{{Alecian}}{2015}]{Alecian15}
{Alecian} G.,  2015, \mn@doi [\mnras] {10.1093/mnras/stv2205}, \href
  {http://adsabs.harvard.edu/abs/2015MNRAS.454.3143A} {454, 3143}

\bibitem[\protect\citeauthoryear{{Alecian} \& {Stift}}{{Alecian} \&
  {Stift}}{2017}]{Alecian:2017aa}
{Alecian} G.,  {Stift} M.~J.,  2017, \mn@doi [\mnras] {10.1093/mnras/stx496},
  \href {http://adsabs.harvard.edu/abs/2017MNRAS.468.1023A} {468, 1023}

\bibitem[\protect\citeauthoryear{{Babcock}}{{Babcock}}{1949}]{Babcock:1949aa}
{Babcock} H.~W.,  1949, \mn@doi [\apj] {10.1086/145192}, \href
  {http://adsabs.harvard.edu/abs/1949ApJ...110..126B} {110, 126}

\bibitem[\protect\citeauthoryear{{Bohlender}, {Landstreet}, {Brown}  \&
  {Thompson}}{{Bohlender} et~al.}{1987}]{Bohlender:1987aa}
{Bohlender} D.~A.,  {Landstreet} J.~D.,  {Brown} D.~N.,   {Thompson} I.~B.,
  1987, \mn@doi [\apj] {10.1086/165830}, \href
  {http://adsabs.harvard.edu/abs/1987ApJ...323..325B} {323, 325}

\bibitem[\protect\citeauthoryear{{Borra} \& {Landstreet}}{{Borra} \&
  {Landstreet}}{1979}]{Borra:1979aa}
{Borra} E.~F.,  {Landstreet} J.~D.,  1979, \mn@doi [\apj] {10.1086/156907},
  \href {http://adsabs.harvard.edu/abs/1979ApJ...228..809B} {228, 809}

\bibitem[\protect\citeauthoryear{{Borra}, {Landstreet}  \& {Thompson}}{{Borra}
  et~al.}{1983}]{Borra:1983aa}
{Borra} E.~F.,  {Landstreet} J.~D.,   {Thompson} I.,  1983, \mn@doi [\apjs]
  {10.1086/190889}, \href {http://adsabs.harvard.edu/abs/1983ApJS...53..151B}
  {53, 151}

\bibitem[\protect\citeauthoryear{{Braithwaite}}{{Braithwaite}}{2007}]{Braithwaite:2007aa}
{Braithwaite} J.,  2007, \mn@doi [\aap] {10.1051/0004-6361:20065903}, \href
  {http://adsabs.harvard.edu/abs/2007A%26A...469..275B} {469, 275}

\bibitem[\protect\citeauthoryear{{Braithwaite}}{{Braithwaite}}{2009}]{Braithwaite:2009aa}
{Braithwaite} J.,  2009, \mn@doi [\mnras] {10.1111/j.1365-2966.2008.14034.x},
  \href {http://adsabs.harvard.edu/abs/2009MNRAS.397..763B} {397, 763}

\bibitem[\protect\citeauthoryear{{Braithwaite} \& {Nordlund}}{{Braithwaite} \&
  {Nordlund}}{2006}]{Braithwaite:2006aa}
{Braithwaite} J.,  {Nordlund} {\AA}.,  2006, \mn@doi [\aap]
  {10.1051/0004-6361:20041980}, \href
  {http://adsabs.harvard.edu/abs/2006A%26A...450.1077B} {450, 1077}

\bibitem[\protect\citeauthoryear{{Cowley}}{{Cowley}}{1972}]{Cowley:1972aa}
{Cowley} A.,  1972, \mn@doi [\aj] {10.1086/111348}, \href
  {http://adsabs.harvard.edu/abs/1972AJ.....77..750C} {77, 750}

\bibitem[\protect\citeauthoryear{{Donati}, {Semel}  \& {Rees}}{{Donati}
  et~al.}{1992}]{Donati:1992aa}
{Donati} J.-F.,  {Semel} M.,   {Rees} D.~E.,  1992, \aap, \href
  {http://adsabs.harvard.edu/abs/1992A%26A...265..669D} {265, 669}

\bibitem[\protect\citeauthoryear{{Donati}, {Semel}, {Carter}, {Rees}  \&
  {Collier Cameron}}{{Donati} et~al.}{1997}]{Donati:1997aa}
{Donati} J.-F.,  {Semel} M.,  {Carter} B.~D.,  {Rees} D.~E.,   {Collier
  Cameron} A.,  1997, \mnras, \href
  {http://adsabs.harvard.edu/abs/1997MNRAS.291..658D} {291, 658}

\bibitem[\protect\citeauthoryear{{Drake}, {Abbott}, {Bastian}, {Bieging},
  {Churchwell}, {Dulk}  \& {Linsky}}{{Drake} et~al.}{1987}]{Drake:1987aa}
{Drake} S.~A.,  {Abbott} D.~C.,  {Bastian} T.~S.,  {Bieging} J.~H.,
  {Churchwell} E.,  {Dulk} G.,   {Linsky} J.~L.,  1987, \mn@doi [\apj]
  {10.1086/165784}, \href {http://adsabs.harvard.edu/abs/1987ApJ...322..902D}
  {322, 902}

\bibitem[\protect\citeauthoryear{{Duez} \& {Mathis}}{{Duez} \&
  {Mathis}}{2010}]{Duez:2010aa}
{Duez} V.,  {Mathis} S.,  2010, \mn@doi [\aap] {10.1051/0004-6361/200913496},
  \href {http://adsabs.harvard.edu/abs/2010A%26A...517A..58D} {517, A58}

\bibitem[\protect\citeauthoryear{{Duez}, {Braithwaite}  \& {Mathis}}{{Duez}
  et~al.}{2010}]{Duez:2010ab}
{Duez} V.,  {Braithwaite} J.,   {Mathis} S.,  2010, \mn@doi [\apjl]
  {10.1088/2041-8205/724/1/L34}, \href
  {http://adsabs.harvard.edu/abs/2010ApJ...724L..34D} {724, L34}

\bibitem[\protect\citeauthoryear{{Edwards}}{{Edwards}}{1932}]{Edwards:1932aa}
{Edwards} D.~L.,  1932, \mn@doi [\mnras] {10.1093/mnras/92.5.389}, \href
  {http://adsabs.harvard.edu/abs/1932MNRAS..92..389E} {92, 389}

\bibitem[\protect\citeauthoryear{{Groote} \& {Hunger}}{{Groote} \&
  {Hunger}}{1976}]{Groote:1976aa}
{Groote} D.,  {Hunger} K.,  1976, \aap, \href
  {http://adsabs.harvard.edu/abs/1976A%26A....52..303G} {52, 303}

\bibitem[\protect\citeauthoryear{{Grunhut} et~al.,}{{Grunhut}
  et~al.}{2017}]{Grunhut:2017aa}
{Grunhut} J.~H.,  et~al., 2017, \mn@doi [\mnras] {10.1093/mnras/stw2743}, \href
  {http://adsabs.harvard.edu/abs/2017MNRAS.465.2432G} {465, 2432}

\bibitem[\protect\citeauthoryear{{Hoffleit} \& {Warren}}{{Hoffleit} \&
  {Warren}}{1995}]{Hoffleit:1995aa}
{Hoffleit} D.,  {Warren} Jr. W.~H.,  1995, VizieR Online Data Catalog, \href
  {http://adsabs.harvard.edu/abs/1995yCat.5050....0H} {5050}

\bibitem[\protect\citeauthoryear{{Kochukhov}}{{Kochukhov}}{2017}]{Kochukhov:2017ab}
{Kochukhov} O.,  2017, \mn@doi [\aap] {10.1051/0004-6361/201629768}, \href
  {http://adsabs.harvard.edu/abs/2017A%26A...597A..58K} {597, A58}

\bibitem[\protect\citeauthoryear{{Kochukhov} \& {Piskunov}}{{Kochukhov} \&
  {Piskunov}}{2002}]{Kochukhov:2002aa}
{Kochukhov} O.,  {Piskunov} N.,  2002, \mn@doi [\aap]
  {10.1051/0004-6361:20020300}, \href
  {http://adsabs.harvard.edu/abs/2002A%26A...388..868K} {388, 868}

\bibitem[\protect\citeauthoryear{{Kochukhov} \& {Wade}}{{Kochukhov} \&
  {Wade}}{2010}]{Kochukhov10}
{Kochukhov} O.,  {Wade} G.~A.,  2010, \mn@doi [\aap]
  {10.1051/0004-6361/200913860}, \href
  {http://adsabs.harvard.edu/abs/2010A%26A...513A..13K} {513, A13}

\bibitem[\protect\citeauthoryear{{Kochukhov} \& {Wade}}{{Kochukhov} \&
  {Wade}}{2016}]{Koch16}
{Kochukhov} O.,  {Wade} G.~A.,  2016, \mn@doi [\aap]
  {10.1051/0004-6361/201527454}, \href
  {http://cdsads.u-strasbg.fr/abs/2016A%26A...586A..30K} {586, A30}

\bibitem[\protect\citeauthoryear{{Kochukhov}, {Makaganiuk}  \&
  {Piskunov}}{{Kochukhov} et~al.}{2010}]{Kochukhov:2010ab}
{Kochukhov} O.,  {Makaganiuk} V.,   {Piskunov} N.,  2010, \mn@doi [\aap]
  {10.1051/0004-6361/201015429}, \href
  {http://adsabs.harvard.edu/abs/2010A%26A...524A...5K} {524, A5}

\bibitem[\protect\citeauthoryear{{Kochukhov}, {Lundin}, {Romanyuk}  \&
  {Kudryavtsev}}{{Kochukhov} et~al.}{2011}]{Kochukhov:2011aa}
{Kochukhov} O.,  {Lundin} A.,  {Romanyuk} I.,   {Kudryavtsev} D.,  2011,
  \mn@doi [\apj] {10.1088/0004-637X/726/1/24}, \href
  {http://adsabs.harvard.edu/abs/2011ApJ...726...24K} {726, 24}

\bibitem[\protect\citeauthoryear{{Kochukhov}, {L{\"u}ftinger}, {Neiner},
  {Alecian}  \& {MiMeS Collaboration}}{{Kochukhov} et~al.}{2014}]{Koch14}
{Kochukhov} O.,  {L{\"u}ftinger} T.,  {Neiner} C.,  {Alecian} E.,   {MiMeS
  Collaboration} 2014, \mn@doi [\aap] {10.1051/0004-6361/201423472}, \href
  {http://adsabs.harvard.edu/abs/2014A%26A...565A..83K} {565, A83}

\bibitem[\protect\citeauthoryear{{Kochukhov} et~al.,}{{Kochukhov}
  et~al.}{2015}]{Kochukhov15}
{Kochukhov} O.,  et~al., 2015, \mn@doi [\aap] {10.1051/0004-6361/201425065},
  \href {http://adsabs.harvard.edu/abs/2015A%26A...574A..79K} {574, A79}

\bibitem[\protect\citeauthoryear{{Kochukhov}, {Silvester}, {Bailey},
  {Landstreet}  \& {Wade}}{{Kochukhov} et~al.}{2017}]{Kochukhov:2017aa}
{Kochukhov} O.,  {Silvester} J.,  {Bailey} J.~D.,  {Landstreet} J.~D.,   {Wade}
  G.~A.,  2017, preprint, \href
  {http://adsabs.harvard.edu/abs/2017arXiv170504966K} {} (\mn@eprint {arXiv}
  {1705.04966})

\bibitem[\protect\citeauthoryear{{Lenz} \& {Breger}}{{Lenz} \&
  {Breger}}{2005}]{Lenz:2005aa}
{Lenz} P.,  {Breger} M.,  2005, \mn@doi [Communications in Asteroseismology]
  {10.1553/cia146s53}, \href
  {http://adsabs.harvard.edu/abs/2005CoAst.146...53L} {146, 53}

\bibitem[\protect\citeauthoryear{{Linsky}, {Drake}  \& {Bastian}}{{Linsky}
  et~al.}{1992}]{Linsky:1992aa}
{Linsky} J.~L.,  {Drake} S.~A.,   {Bastian} T.~S.,  1992, \mn@doi [\apj]
  {10.1086/171509}, \href {http://adsabs.harvard.edu/abs/1992ApJ...393..341L}
  {393, 341}

\bibitem[\protect\citeauthoryear{{Markey} \& {Tayler}}{{Markey} \&
  {Tayler}}{1973}]{Markey:1973aa}
{Markey} P.,  {Tayler} R.~J.,  1973, \mn@doi [\mnras] {10.1093/mnras/163.1.77},
  \href {http://adsabs.harvard.edu/abs/1973MNRAS.163...77M} {163, 77}

\bibitem[\protect\citeauthoryear{{Mihalas} \& {Henshaw}}{{Mihalas} \&
  {Henshaw}}{1966}]{Mihalas:1966aa}
{Mihalas} D.,  {Henshaw} J.~L.,  1966, \mn@doi [\apj] {10.1086/148588}, \href
  {http://adsabs.harvard.edu/abs/1966ApJ...144...25M} {144, 25}

\bibitem[\protect\citeauthoryear{{Molnar}}{{Molnar}}{1972}]{Molnar:1972aa}
{Molnar} M.~R.,  1972, \mn@doi [\apj] {10.1086/151570}, \href
  {http://adsabs.harvard.edu/abs/1972ApJ...175..453M} {175, 453}

\bibitem[\protect\citeauthoryear{{Piskunov} \& {Kochukhov}}{{Piskunov} \&
  {Kochukhov}}{2002}]{Piskunov:2002aa}
{Piskunov} N.,  {Kochukhov} O.,  2002, \mn@doi [\aap]
  {10.1051/0004-6361:20011517}, \href
  {http://adsabs.harvard.edu/abs/2002A%26A...381..736P} {381, 736}

\bibitem[\protect\citeauthoryear{{Piskunov}, {Kupka}, {Ryabchikova}, {Weiss}
  \& {Jeffery}}{{Piskunov} et~al.}{1995}]{Piskunov:1995aa}
{Piskunov} N.~E.,  {Kupka} F.,  {Ryabchikova} T.~A.,  {Weiss} W.~W.,
  {Jeffery} C.~S.,  1995, \aaps, \href
  {http://adsabs.harvard.edu/abs/1995A%26AS..112..525P} {112, 525}

\bibitem[\protect\citeauthoryear{{Ros{\'e}n}, {Kochukhov}  \&
  {Wade}}{{Ros{\'e}n} et~al.}{2015}]{Rosen15}
{Ros{\'e}n} L.,  {Kochukhov} O.,   {Wade} G.~A.,  2015, \mn@doi [\apj]
  {10.1088/0004-637X/805/2/169}, \href
  {http://adsabs.harvard.edu/abs/2015ApJ...805..169R} {805, 169}

\bibitem[\protect\citeauthoryear{{Rusomarov}, {Kochukhov}, {Ryabchikova}  \&
  {Piskunov}}{{Rusomarov} et~al.}{2015}]{Rusomarov15}
{Rusomarov} N.,  {Kochukhov} O.,  {Ryabchikova} T.,   {Piskunov} N.,  2015,
  \aap, \href {http://adsabs.harvard.edu/abs/2014arXiv1409.6955R} {573, A123}

\bibitem[\protect\citeauthoryear{{Rusomarov}, {Kochukhov}, {Ryabchikova}  \&
  {Ilyin}}{{Rusomarov} et~al.}{2016}]{Rusomarov16}
{Rusomarov} N.,  {Kochukhov} O.,  {Ryabchikova} T.,   {Ilyin} I.,  2016,
  \mn@doi [\aap] {10.1051/0004-6361/201527719}, \href
  {http://adsabs.harvard.edu/abs/2016A%26A...588A.138R} {588, A138}

\bibitem[\protect\citeauthoryear{{Ryabchikova} \& {Stateva}}{{Ryabchikova} \&
  {Stateva}}{1996}]{Ryabchikova:1996aa}
{Ryabchikova} T.~A.,  {Stateva} I.,  1996, in {Adelman} S.~J.,  {Kupka} F.,
  {Weiss} W.~W.,  eds,  Astronomical Society of the Pacific Conference Series
  Vol. 108, M.A.S.S., Model Atmospheres and Spectrum Synthesis. p.~265

\bibitem[\protect\citeauthoryear{{Ryabchikova}, {Piskunov}, {Kurucz},
  {Stempels}, {Heiter}, {Pakhomov}  \& {Barklem}}{{Ryabchikova}
  et~al.}{2015}]{Ryabchikova:2015aa}
{Ryabchikova} T.,  {Piskunov} N.,  {Kurucz} R.~L.,  {Stempels} H.~C.,  {Heiter}
  U.,  {Pakhomov} Y.,   {Barklem} P.~S.,  2015, \mn@doi [\physscr]
  {10.1088/0031-8949/90/5/054005}, \href
  {http://adsabs.harvard.edu/abs/2015PhyS...90e4005R} {90, 054005}

\bibitem[\protect\citeauthoryear{{Sadakane}}{{Sadakane}}{1984}]{Sadakane:1984aa}
{Sadakane} K.,  1984, \mn@doi [\pasp] {10.1086/131330}, \href
  {http://adsabs.harvard.edu/abs/1984PASP...96..259S} {96, 259}

\bibitem[\protect\citeauthoryear{{Sargent}, {Greenstein}  \&
  {Sargent}}{{Sargent} et~al.}{1969}]{Sargent:1969aa}
{Sargent} A.~I.,  {Greenstein} J.~L.,   {Sargent} W.~L.~W.,  1969, \mn@doi
  [\apj] {10.1086/150111}, \href
  {http://adsabs.harvard.edu/abs/1969ApJ...157..757S} {157, 757}

\bibitem[\protect\citeauthoryear{{Searle} \& {Sargent}}{{Searle} \&
  {Sargent}}{1964}]{Searle:1964aa}
{Searle} L.,  {Sargent} W.~L.~W.,  1964, \mn@doi [\apj] {10.1086/147817}, \href
  {http://adsabs.harvard.edu/abs/1964ApJ...139..793S} {139, 793}

\bibitem[\protect\citeauthoryear{{Shore}, {Brown}  \& {Sonneborn}}{{Shore}
  et~al.}{1987}]{Shore:1987aa}
{Shore} S.~N.,  {Brown} D.~N.,   {Sonneborn} G.,  1987, \mn@doi [\aj]
  {10.1086/114512}, \href {http://adsabs.harvard.edu/abs/1987AJ.....94..737S}
  {94, 737}

\bibitem[\protect\citeauthoryear{{Shore}, {Brown}, {Sonneborn}, {Landstreet}
  \& {Bohlender}}{{Shore} et~al.}{1990}]{Shore:1990aa}
{Shore} S.~N.,  {Brown} D.~N.,  {Sonneborn} G.,  {Landstreet} J.~D.,
  {Bohlender} D.~A.,  1990, \mn@doi [\apj] {10.1086/168233}, \href
  {http://adsabs.harvard.edu/abs/1990ApJ...348..242S} {348, 242}

\bibitem[\protect\citeauthoryear{{Silvester}, {Kochukhov}  \&
  {Wade}}{{Silvester} et~al.}{2014a}]{Silvester14a}
{Silvester} J.,  {Kochukhov} O.,   {Wade} G.~A.,  2014a, \mn@doi [\mnras]
  {10.1093/mnras/stu306}, \href
  {http://adsabs.harvard.edu/abs/2014MNRAS.440..182S} {440, 182}

\bibitem[\protect\citeauthoryear{{Silvester}, {Kochukhov}  \&
  {Wade}}{{Silvester} et~al.}{2014b}]{Silvester14b}
{Silvester} J.,  {Kochukhov} O.,   {Wade} G.~A.,  2014b, \mn@doi [\mnras]
  {10.1093/mnras/stu1531}, \href
  {http://adsabs.harvard.edu/abs/2014MNRAS.444.1442S} {444, 1442}

\bibitem[\protect\citeauthoryear{{Silvester}, {Kochukhov}  \&
  {Wade}}{{Silvester} et~al.}{2015}]{Silvester15}
{Silvester} J.,  {Kochukhov} O.,   {Wade} G.~A.,  2015, \mn@doi [\mnras]
  {10.1093/mnras/stv1775}, \href
  {http://adsabs.harvard.edu/abs/2015MNRAS.453.2163S} {453, 2163}

\bibitem[\protect\citeauthoryear{{Smith}, {Wade}, {Bohlender}  \&
  {Bolton}}{{Smith} et~al.}{2006}]{Smith:2006aa}
{Smith} M.~A.,  {Wade} G.~A.,  {Bohlender} D.~A.,   {Bolton} C.~T.,  2006,
  \mn@doi [\aap] {10.1051/0004-6361:20054760}, \href
  {http://adsabs.harvard.edu/abs/2006A%26A...458..581S} {458, 581}

\bibitem[\protect\citeauthoryear{{Stibbs}}{{Stibbs}}{1950}]{Stibbs:1950aa}
{Stibbs} D.~W.~N.,  1950, \mnras, \href
  {http://adsabs.harvard.edu/abs/1950MNRAS.110..395S} {110, 395}

\bibitem[\protect\citeauthoryear{{Stift} \& {Alecian}}{{Stift} \&
  {Alecian}}{2016}]{Stift:2016aa}
{Stift} M.~J.,  {Alecian} G.,  2016, \mn@doi [\mnras] {10.1093/mnras/stv2962},
  \href {http://adsabs.harvard.edu/abs/2016MNRAS.457...74S} {457, 74}

\bibitem[\protect\citeauthoryear{{Wade}, {Donati}, {Landstreet}  \&
  {Shorlin}}{{Wade} et~al.}{2000}]{Wade:2000ab}
{Wade} G.~A.,  {Donati} J.-F.,  {Landstreet} J.~D.,   {Shorlin} S.~L.~S.,
  2000, \mn@doi [\mnras] {10.1046/j.1365-8711.2000.03273.x}, \href
  {http://adsabs.harvard.edu/abs/2000MNRAS.313..823W} {313, 823}

\bibitem[\protect\citeauthoryear{{Wade} et~al.,}{{Wade}
  et~al.}{2006}]{Wade:2006aa}
{Wade} G.~A.,  et~al., 2006, \mn@doi [\aap] {10.1051/0004-6361:20054759}, \href
  {http://adsabs.harvard.edu/abs/2006A%26A...458..569W} {458, 569}

\bibitem[\protect\citeauthoryear{{Walborn}}{{Walborn}}{1974}]{Walborn:1974aa}
{Walborn} N.~R.,  1974, \mn@doi [\apjl] {10.1086/181558}, \href
  {http://adsabs.harvard.edu/abs/1974ApJ...191L..95W} {191, L95}

\bibitem[\protect\citeauthoryear{{Yakunin} et~al.,}{{Yakunin}
  et~al.}{2015}]{Yakunin:2015aa}
{Yakunin} I.,  et~al., 2015, \mn@doi [\mnras] {10.1093/mnras/stu2401}, \href
  {http://adsabs.harvard.edu/abs/2015MNRAS.447.1418Y} {447, 1418}

\bibitem[\protect\citeauthoryear{{Zima}}{{Zima}}{2008}]{Zima:2008aa}
{Zima} W.,  2008, \mn@doi [Communications in Asteroseismology]
  {10.1553/cia155s17}, \href
  {http://adsabs.harvard.edu/abs/2008CoAst.155...17Z} {155, 17}

\makeatother
\end{thebibliography}








\bsp	
\label{lastpage}
\end{document}